# End-to-end design of multicolor scintillators for enhanced energy resolution in X-ray imaging

Seokhwan Min[1,2], Seou Choi[1], Simo Pajovic[3], Sachin Vaidya[4], Nicholas Rivera[5], Shanhui Fan[6], Marin Soljačić[1,4], Charles Roques-Carmes[1,6*]

[1]Research Laboratory of Electronics, Massachusetts Institute of Technology, 77 Massachusetts Ave., Cambridge, 02139, MA, USA.
[2]Department of Material Science and Engineering, Korea Advanced Institute of Science and Technology, 291 Daehak-ro, Daejeon, 34141, Daejeon, Republic of Korea.
[3]Department of Mechanical Engineering, Massachusetts Institute of Technology, 77 Massachusetts Ave., Cambridge, 02139, MA, USA.
[4]Department of Physics, Massachusetts Institute of Technology, 77 Massachusetts Ave., Cambridge, 02139, MA, USA.
[5]Department of Physics, Harvard University, Massachusetts Hall, Cambridge, 02138, MA, USA.
[6]*E. L. Ginzton Laboratories, Stanford University, 450 Jane Stanford Way, Stanford, 94305-2048, CA, USA.

*Corresponding author(s). E-mail(s): chrc@stanford.edu; Tel. +1-650-723-2300
Contributing authors: seokhwan.min@kaist.ac.kr; seouc130@mit.edu; pajovics@mit.edu; svaidya1@mit.edu; nrivera@fas.harvard.edu; shanhui@stanford.edu; soljacic@mit.edu;

# Abstract

Scintillators have been widely used in X-ray imaging due to their ability to convert high-energy radiation into visible light, making them essential for applications such as medical imaging and high-energy physics. Recent advances in the artificial structuring of scintillators offer new opportunities for improving the energy resolution of scintillator-based X-ray detectors. Here, we present a three-bin energy-resolved X-ray imaging framework based on a three-layer multicolor scintillator used in conjunction with a physics-aware image postprocessing algorithm. The multicolor scintillator is able to preserve X-ray energy information through the combination of emission wavelength multiplexing and energy-dependent isolation of X-ray absorption in specific layers. The dominant emission color and the radius of the spot measured by the detector are used to infer the incident X-ray energy based on prior knowledge of the energy-dependent absorption profiles of the scintillator stack. Through *ab initio* Monte Carlo simulations, we show that our approach can achieve an energy reconstruction accuracy of 49.7%, which is only 2% below the maximum accuracy achievable with realistic scintillators. We apply our framework to medical phantom imaging simulations where we demonstrate that it can effectively differentiate iodine and gadolinium-based contrast agents from bone, muscle, and soft tissue.

# 1. Introduction

Scintillators are an effective means of converting high-energy particles and radiation into visible light. Therefore, they have found widespread use in a multitude of applications such as high-energy particle detection and medical imaging techniques such as X-ray radiography and computed tomography (CT)[1,2]. Much effort has been put into improving the image quality of scintillation detectors for these applications. Early research mainly focused on the development of new scintillating materials, which saw the development of dense, fast, and high-light-yield scintillators such as Gadox, LSO:Ce, and LYSO:Ce[1,3,4]. On the other hand, nanophotonic and multilayer engineering of scintillators have recently emerged as promising alternative approaches to improve the sensitivity and resolution of scintillation-based imaging[5–7]. These studies indicate that the artificial structuring of scintillators can provide benefits that are significant and complement those achievable with material innovations.

Artificial structuring can be used not only to shape the scintillation emission, but also to encode X-ray information into the emitted photons. One significant challenge in X-ray imaging remains the ability to accurately resolve the energy of absorbed X-rays, which is crucial for material differentiation in medical diagnostics. Therefore, much effort has been put into developing techniques that enable energy-resolved X-ray imaging, such as K-edge subtraction[8], dual source imaging[9,10], and more recently, photon-counting detectors[11,12]. These methods suffer from some of the following constraints: (1) monochromatic X-ray requirement incompatible with clinical settings (K-edge subtraction); (2) considerable hardware modifications or longer image acquisitions from switching source voltages (dual source imaging); (3) limitations in pixel sizes due to required electronic circuitry (photon-counting detectors); and (4) high cost. Recent developments in low-cost synthesis of lead-halide perovskites for direct X-ray detection have demonstrated promising steps towards addressing these limitations[13].

Here, we present a framework for energy-resolved X-ray imaging based on multilayer, multicolor (each layer emitting a different color) scintillators – being fully compatible with conventional X-ray tubes and detectors, avoiding the technical difficulties of developing high-flux, physically compact monochromatic sources, extensive hardware modifications, or complex acquisition setups. These scintillators are designed in conjunction with a physics-aware clustering algorithm for the postprocessing of the color images obtained using the multicolor scintillator. We develop a Monte Carlo simulation framework to design a multicolor three-layer scintillator stack achieving near-optimal energy reconstruction accuracy for enhanced X-ray energy discrimination and improved medical imaging, with further performance gains possible using nanophotonic structures. This approach paves the way for advanced X-ray detectors with superior energy resolution, with potential applications in diagnostic imaging and material identification.

Specifically, we develop our framework with *ab initio* Monte Carlo simulations using real scintillator material parameters. We propose a design for a three-layer scintillator stack whose layers emit red, green, and blue scintillation photons, respectively (Fig. 1) such that the colors can be easily distinguished with commercial red-green-blue (RGB) detectors. The constituent scintillators and their thicknesses are designed for three-bin energy discrimination in the energy range of 16-67 keV, which is tailored for the identification of iodine and gadolinium-based contrast agents with absorption K-edges at 33 and 50 keV. We benchmark the performance of our framework on specific tasks including X-ray energy reconstruction and medical phantom imaging. The multicolor scintillator was able to achieve energy reconstruction accuracy of up to 49.7%, which is within 2% of the accuracy upper bound achievable using scintillators with realistic absorption coefficients. The accuracy is 5% higher than that of a ZnSe:Te single-color scintillator with the same overall thickness. This is a significant improvement given that the accuracy upper ($\approx$ 52%) and lower ($\approx$ 33.3%) bounds span only a range of $\approx$ 19%. Medical phantom simulations indicate that the increased energy accuracy of the multicolor scintillator results in a noticeable improvement in the ability of our postprocessing algorithm to discriminate between contrast agents and various tissues. Finally, we show that nanophotonic structures may be used to further improve the performance of multilayer scintillators. Namely, we show that a one-dimensional nanophotonic filter can be used to increase the X-ray photon flux density that can be processed by our framework.

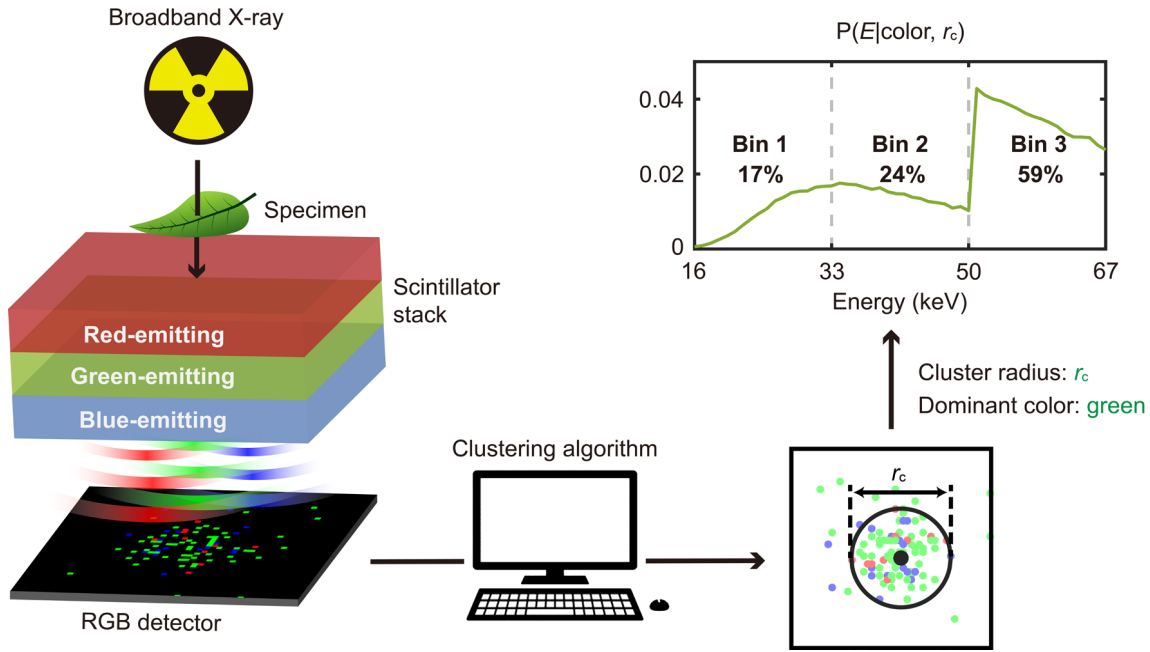

**Fig. 1 Schematic of the energy reconstruction process using a multicolor scintillator.** X-rays from a typical broadband source (X-ray tube) are incident on the imaged specimen. The photons that pass through are absorbed in various depths of the multicolor scintillator, whose emissions create colored spots on the red-green-blue (RGB) detector. The detector image is fed into a clustering algorithm which identifies each cluster's position, radius, and dominant color. This information is used to obtain the probability that the cluster was created by an X-ray photon from a given energy bin.

Compared to previous methods that were based on maximizing the amount of information using either multiple X-ray sources or photon-counting detectors, multilayer multicolor scintillators consist of a completely different approach that focuses on the efficient transfer of energy information from X-rays to optical photons. Scintillation photons are potential sources of a rich variety of information encoded through properties such as wavelength, amplitude, polarization, propagation direction, and emission pulse width among many others. In particular, wavelength multiplexing of scintillation emission using multiple different scintillators naturally provides valuable information about the X-ray photon absorption location, and is relatively straightforward to implement using heterostructures, e.g., by bonding together bulk scintillators. Due to the fact that X-rays have spatial absorption profiles that vary strongly with energy, wavelength multiplexing with multicolor scintillators is an effective means of encoding and extracting energy information for X-ray imaging[14,15]. Our methods are also a first example of end-to-end[16,17] design of scintillators with optimized (nanophotonic) structural properties: the optimized scintillator acts as a "physical preprocessor" for the optical signal, input to a reconstruction algorithm that then identifies properties of the incident X-rays. The "pre-processor" digests the raw data (incoming X-ray energies) into a form that the post-processor can interpret effectively and the post-processor extracts useful information out of the otherwise unintelligible "pre-processed" data. This method of design has been effectively used in image processing in particular to reconstruct at high resolution a much greater amount of spatial, spectral, and polarization information compared to post-processing-only systems.

## 2. Results

### 2.1 Working Principles of the Multicolor Scintillator

The X-ray absorption profile along the depth of any medium, including scintillators, varies strongly with the incident energy. This is a straightforward consequence of Beer's Law, which allows the absorption per unit length to be written as follows:

$$A(z, E) = -\frac{dT(z,E)}{dz} = \mu(z, E) \exp[-\mu(z, E)z] \quad (1)$$

where $z$ is the depth measured from the X-ray impact position on the scintillator, $E$ is the X-ray energy, and $\mu$ is the linear attenuation coefficient, which can be depth-dependent in a multilayer stack. $T$ represents the transmitted proportion of the X-rays with a given energy at a given depth. The absorption profile, therefore, follows an exponential curve with respect to $z$, where the attenuation strength is dependent on $\mu$. In the absence of absorption edges, linear attenuation coefficients monotonically decrease with increasing energy in the clinical X-ray energy range. This means low-energy absorption will be highly confined to the upper region of the scintillator, whereas at higher energies, the absorption will spread out more evenly across the entirety of the scintillator. The absorption profile expressed as a probability density function over $z$ is:

$$p_{abs}(z, E_{inc}) = \frac{A(z, E_{inc})}{\int_0^L A(z', E_{inc}) dz'} \quad (2)$$

where $E_{inc}$ is the incident energy and $L$ is the overall thickness of the scintillator stack. Thus, given $N(E_i)\, dE_{inc}$ incident X-ray photons with energy $E_i \leq E_{inc} \leq E_i + dE_{inc}$, an amount of $N(E_i)\, p_{abs}(z_p, E_i)\, dE_{inc}\, dz$ will be absorbed between $z = z_p$ and $z = z_p + dz$.

  The postprocessing algorithm has no knowledge of $E_i$ and therefore must rely on estimating $E_i$ based on the absorption depths of the X-ray photons. Assuming that there is an accurate method for estimating the absorption depth of each X-ray photon from the detector measurements, the absorbed X-ray energy distribution at a depth $z$ as a probability density function over $E$:

$$p_{\text{det}}(z, E_{\text{det}}) = \frac{A(z, E_{\text{det}})}{\int_{E_{\text{low}}}^{E_{\text{high}}} A(z, E_{\text{det}}) dE_{\text{det}}} \quad (3)$$

where $E_{\text{det}}$ is the energy measured at the detector. $E_{\text{low}}$ and $E_{\text{high}}$ represent either ends of the target energy range. Eq. (3) is used by the postprocessing algorithm to guess the incident energy in a probabilistic manner. Therefore, the amount of incident X-ray photons with energy $E_i \leq E_{\text{inc}} \leq E_i + dE_{\text{inc}}$ that is absorbed at $z_p \leq z \leq z_p + dz$ and then assigned a reconstructed energy value of $E_j \leq E_{\text{det}} \leq E_j + dE_{\text{det}}$ is $N(E_i) \, p_{\text{abs}}(z_p, E_i) \, p_{\text{det}}(z_p, E_j)$ $dE_{\text{inc}} \, dz \, dE_{\text{det}}$. X-ray photons can take different paths (e.g., get absorbed at different depths) to eventually be assigned an energy of $E_j \leq E_{\text{det}} \leq E_j + dE_{\text{det}}$. Therefore, the probability density function that maps the incident energy $E_{\text{inc}}$ to the reconstructed energy $E_{\text{det}}$ is:

$$p_{\text{reconstr}}(E_{\text{inc}}, E_{\text{det}}) dE_{\text{inc}} \, dE_{\text{det}} = \frac{\int_0^L N(E_{\text{inc}}) p_{\text{abs}}(z, E_{\text{inc}}) p_{\text{det}}(z, E_{\text{det}}) dE_{\text{inc}} \, dz \, dE_{\text{det}}}{\int_{E_{\text{low}}}^{E_{\text{high}}} A(z, E_{\text{det}}) dE_{\text{det}}} \quad (4)$$

When the energy range is divided into discrete bins, the proportion of X-rays from bin $m$ $\left(E_1^{(m)} \leq E \leq E_2^{(m)}\right)$ that is reconstructed as being in bin $n$ $\left(E_1^{(n)} \leq E \leq E_2^{(n)}\right)$ is given by:

$$p_{\text{reconstr}}^{(mn)} = \int_{E_1^{(n)}}^{E_2^{(n)}} \int_{E_1^{(m)}}^{E_2^{(m)}} p_{\text{reconstr}}(E_{\text{inc}}, E_{\text{det}}) dE_{\text{inc}} \, dE_{\text{det}} \quad (5)$$

From this, one can define the following energy reconstruction accuracy metric which measures the average proportion of the incident X-ray photons that is classified into the ground truth energy bin:

$$\eta = \sum_{m=1}^{M} p_{\text{reconstr}}^{(mn)} \qquad (6)$$

where *M* is the total number of energy bins. $\eta$ ranges from 0 to 1 with 1 indicating that all X-ray photons were correctly classified.

This method of energy inference is applicable regardless of the specific structure of the scintillator. Even in single-layer scintillators, the proportion of higher-energy X-rays increases with the absorption depth, which allows some degree of energy discrimination. However, the scintillators need to be thick (> 1 mm) in order to fully absorb high-energy X-rays and achieve good energy reconstruction accuracy, which degrades the image resolution[6] (Fig. S1). Furthermore, single-layer scintillators with an absorption edge within the target energy range have even worse energy accuracy due to the additional ambiguity in $p_{\text{det}}(z,E)$ caused by the absorption edge (Fig. S2).

Instead, we outline the following design principles for reducing ambiguity in energy reconstruction, relying on a three-layer scintillator stack where each layer absorbs a specific energy bin within the target energy range (16-67 keV). The energy range was chosen such that it encompasses the K-edge energies of both I (iodine) and Gd (gadolinium) found in X-ray contrast agents. It was divided into three energy bins based on the K-edge energies of I and Gd (33 and 50 keV). As further discussed below, the placement of energy bin boundaries at the K-edge energies is crucial for engineering drastically different absorption profiles for each energy bin, which is necessary for isolating the energy bins' absorption to specific layers.

Due to the exponential nature of the absorption profile, the absorption of low-energy X-rays is most easily confined to the topmost layer. In order to further facilitate the confinement, the linear attenuation coefficient ($\mu$) of the top scintillator must be large in the lowest energy bin (16-33 keV) and decrease monotonically across the other bins (33-67 keV) without any K-edges (Fig. 2a). The presence of a K-edge in the top layer, at energy above the lowest energy bin, introduces uncertainty as to whether an X-ray photon absorbed in the top layer was in the lowest energy bin or a higher energy bin and should be avoided.

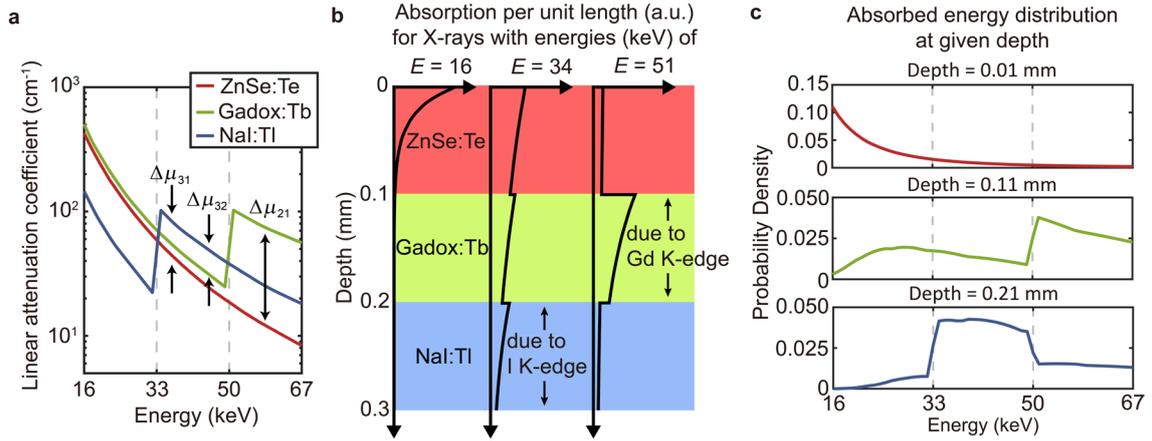

**Fig. 2 Working principle of the energy-resolving multicolor scintillator. a** Linear attenuation coefficients of the different scintillators consisting the multilayer stack (ZnSe:Te, Gadox:Tb, NaI:Tl). **b** Absorption per unit length along the X-ray propagation path for different X-ray energies. The absorption of each energy bin is confined as much as possible to the target layer for each bin (low energies to ZnSe:Te, intermediate energies to NaI:Tl, high energies to Gadox:Tb). **c** Absorbed energy distributions at the top of each layer.

The middle layer is designed to absorb high-energy X-rays (50-67 keV) that pass through the top layer while simultaneously transmitting X-rays in the intermediate energy bin. This can be achieved by using a scintillator that contains Gd (gadolinium) and therefore has a K-edge at 50 keV. In order to ensure that only a minimal proportion of high-energy X-rays are stopped by the top layer, it is important to maximize $\mu_2 d_2 - \mu_1 d_1$ where $d$ is the layer thickness and the subscripts indicate the layer number indexed from the top. In practice, it is more convenient to maximize $\Delta\mu_{21}$ as shown in Fig. 2a, because varying $d$ also affects the absorption of all of the other energy bins. For instance, decreasing $d_1$ would increase the isolation of the highest energy bin in the middle layer, but it would also decrease the isolation of the lowest energy bin in the top layer.

The remaining X-rays with intermediate energies (33-50 keV) are absorbed by the bottom layer. In principle, the bottom layer does not need to transmit any of the energy bins and so does not need to have a K-edge. Nevertheless, in order to minimize the overall thickness of the scintillator stack, it is beneficial for the scintillator to contain I (iodine) with a K-edge at 33 keV such that $\Delta\mu_{31}$ and $\Delta\mu_{32}$ are maximized in the intermediate energy bin (Fig. 2a). Finally, in order to minimize photoluminescent self-absorption between the different layers as the photons propagate towards the detector, the scintillation spectrum should blueshift from top to bottom.

As shown in Fig. 2a, ZnSe:Te, Gadox:Tb, and NaI:Tl were selected as the scintillators that best satisfied the above attenuation coefficient and emission spectrum requirements. Fig. 2b shows the absorption profiles across the scintillator depth calculated using Eq. (1) for specific energies from each energy bin (each layer being 100 μm thick). The confinement of the low and high energy bins to the top and middle layers can be clearly seen. Due to the limited choice of scintillator materials, $\Delta\mu_{31}$ and $\Delta\mu_{32}$ could not be made large enough to improve the confinement of intermediate energies to the bottom layer beyond what is shown in the figure. Nevertheless, the conditional probabilities (Eq. (3)) computed at the top of each layer are maximized at their respective target energy bins as intended. Therefore, the absorption depth provides meaningful information about the energy of the X-ray photon (Fig. 2c).

However, in actual implementation, the only directly measurable information comes from the scintillation spots measured on the RGB detector. Three different types of information can be extracted: lateral position, color, and radius. The lateral position directly indicates the lateral position at which the X-ray was absorbed, which is straightforward. The color and radius are indicators of the absorption depth, but both quantities have associated uncertainties.

Although the scintillation color is a straightforward indicator of the *layer* in which the X-ray was absorbed, it does not provide any information on the exact absorption depth *within* the layer. As seen in Fig. 2b, the absorption profiles not only vary between layers but also within each layer, and this extra information can be used to improve the energy accuracy of the multicolor scintillator.

On the other hand, since scintillation emission is isotropic, the cluster size is directly related to the distance between the point of emission and the detector. However, it is also not suitable as the sole metric for the absorption depth. Typical scintillation spots do not have clear boundaries that can be used to determine their sizes. Instead, we defined an effective cluster radius as the average Euclidean distance from all points in the cluster to the cluster centroid. This metric is robust to outliers but depends on scintillation photon statistics (i.e., scintillation from the same depth can result in a range of cluster radii). This is especially noticeable when there are relatively few photons per cluster, as is the case for scintillators in the energy range of our interest (Fig. S4). The effect of this uncertainty is that absorption profiles inferred using cluster radii are "blurred" versions of the actual absorption profiles (Fig. S5). The addition of color information can help distinguish the layer in which the X-ray was absorbed near layer boundaries, improving energy accuracy (Fig. S6).

Therefore, our postprocessing algorithm approximates the true depth-dependent energy distribution with a color-and-radius-dependent energy distribution in order to determine the X-ray energy bin.

## 2.2 Energy and Image Reconstruction

To empirically obtain the energy distribution data as a function of the cluster color and radius, $10^5$ single X-ray incidence simulations were performed for energies between 16 keV and 67 keV in 1 keV intervals. The simulations were done using GEANT4[18], a Monte Carlo simulation software for high-energy particles (more details may be found in Methods). The energy reconstruction for multicolor and single-color scintillators were then computed using an adapted version of Eqs. (4-6), where $z$ was replaced by the cluster radius $R_c$ and dominant color (or only $R_c$ for single-color scintillators). The layer thicknesses were optimized under different upper bound constraints using a simplified and differentiable surrogate model (see Supplementary Information, Section S3).

As can be seen from Fig. 3a, the multicolor scintillator has the highest accuracy, followed by ZnSe:Te and NaI:Tl. This validates our approach of isolating the absorption of each energy bin to specific layers as described in Section 2.1. As mentioned earlier, the reason for the drastic difference between the accuracies for ZnSe:Te and NaI:Tl is due to the absorption K-edge of NaI:Tl. There is also a general trend of increasing accuracy with overall thickness, which is due to the fact that longer propagation distances lead to better separation of higher energies from intermediate energies. Once the scintillator reaches a certain thickness, the majority of incident X-rays are absorbed in the upper to intermediate depths of the scintillator, and increasing the thickness further will not have any effect on the absorption profile and energy accuracy. For the multicolor scintillator, the portion of the incident X-rays absorbed by the ZnSe:Te layer increases with the thicknesses of each layer. Therefore, at large overall thicknesses, its energy accuracy asymptotically converges to the energy accuracy of the ZnSe:Te single-color scintillator.

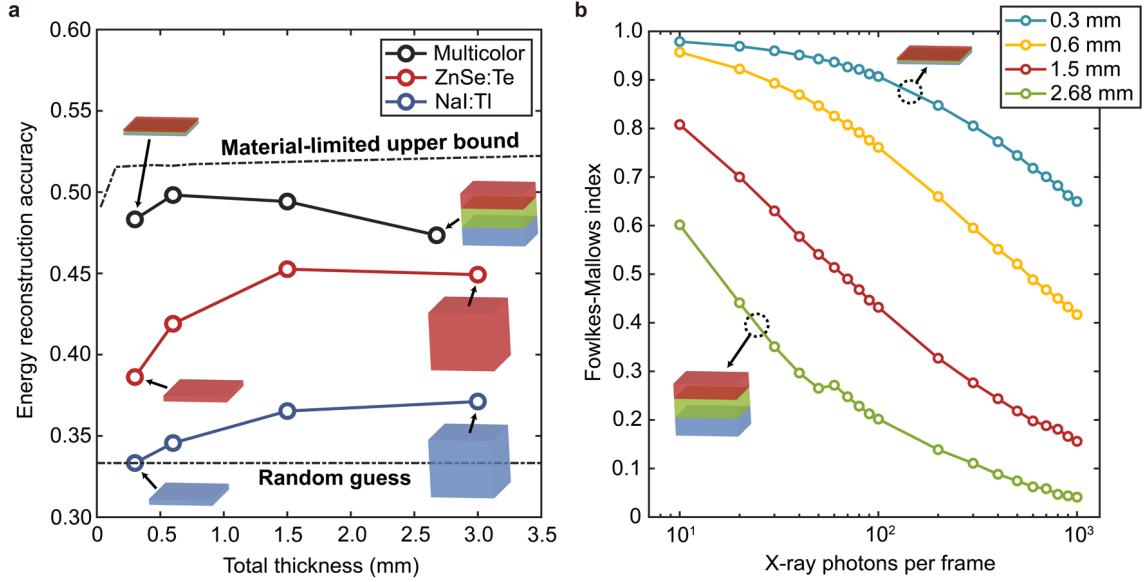

**Fig. 3 Energy and image reconstruction accuracy metrics. a** Energy classification accuracy for the multicolor, ZnSe:Te, and NaI:Tl scintillators of various thicknesses. The reconstruction accuracy is defined as the proportion of X-rays classified into the correct energy bin. **b** Fowlkes-Mallows index for multicolor scintillators of various thicknesses plotted against the average number of X-ray photons incident per frame. The Fowlkes-Mallows index is a measure of the similarity of two clustering results. Here, the output of the clustering algorithm is compared with the ground truth clustering, meaning higher index values correspond to greater clustering accuracy.

For reference, lower and upper bounds on the energy accuracy are plotted with dashed lines in Fig. 3a. Details on the computation of the bounds can be found in Methods. Within the range of possible accuracy values between these bounds, the improvement provided by the multicolor scintillator over the single-color scintillator is non-negligible. This is especially pronounced at lower overall thicknesses, which correspond to the thicknesses of scintillators used in commercial X-ray imaging systems (0.15–0.6 mm)[19,20]. Furthermore, the accuracy of the multicolor scintillator is at best within 2% of the upper bound, which indicates that the specific scintillators selected for each layer are close to optimal.

The energy accuracies in Fig. 3a were computed assuming only a single X-ray photon is incident on the scintillator per detector frame. In practice, much larger numbers of X-rays are incident per frame at different lateral positions, which may cause overlapping of the scintillation photon clusters. Since we used a modified *k*-means clustering algorithm to identify each cluster, overlapping necessarily introduces errors in the cluster radius estimation. This puts an upper limit to the X-ray photon flux density that our energy reconstruction framework can handle without significant degradation of energy and spatial resolution. The Fowlkes-Mallows (FM) index (Eq. (7)) is a measure of similarity between two clustering results, and when one of the results being compared is the ground truth clustering, it can act as a measure of clustering accuracy[21]:

$$\text{FM} = \sqrt{\frac{\text{TP}}{\text{TP+FP}} \cdot \frac{\text{TP}}{\text{TP+FN}}} \qquad (7)$$

where TP is the number of pairs of points that are in the same cluster in both clustering results (true positives). FP and FN represent the number of pairs that are in the same cluster in only one of the clustering results (false positives/negatives). Values of FM range from 0 (complete mismatch) to 1 (exact same clustering).

To investigate the trade-off between the X-ray flux density and clustering accuracy, we simulated the X-ray detection and postprocessing under multiple X-ray photon incidence. A detector area of 1.28×1.28 cm$^2$ was used in the simulations. It is well-known that thinner scintillators have better image resolution due to reduced spread of emissions, and indeed Fig. 3b shows that reducing the thickness improves clustering accuracy (data for single-color scintillators show very similar trends (Fig. S3). In particular, the 0.3 mm scintillator maintains good accuracy (FM ≥ 0.8) up to ≈ 300 X-ray photons per frame. The X-ray flux corresponding to this value depends on the frame rate of the RGB detector being used. Commercially available high-end, high-speed cameras can reach frame rates of up to $10^9$ at 1000 × 860 pixel image resolution[22], which translates to an X-ray flux of $\approx 10^{11} \text{cm}^{-2}\text{s}^{-1}$ commonly encountered in medical imaging. Even with cameras with lower frame rates between $\approx 10^4$ and $\approx 10^5$ (ref. 23), the 0.3 mm scintillator can still handle a flux of $\approx 10^6$ to $\approx 10^7$ $\text{cm}^{-2}\text{s}^{-1}$, which can be meaningful for other applications such as radiotherapy imaging and diagnosis[24].

## 2.3 Material Identification from Reconstructed Images

As mentioned earlier, the specific choice of the energy bins and scintillators in our work is tailored for the identification of iodine and gadolinium-based contrast agents from biological tissue. To demonstrate this, we simulated the imaging and postprocessing steps of a two-dimensional medical phantom. The phantom is 1 cm thick and consists of four different cylindrical regions embedded in a background of adipose tissue (Fig. 4a). The cylindrical regions consists of mineral bone, muscle tissue, 0.9% iodinated blood, and 0.5% gadolinated blood, along with certain proportions of adipose tissue as indicated in Fig. 4a. The iodine and gadolinium concentrations are based on medically relevant values reported in the literature[25]. The phantom was designed such that the cylindrical regions all have the same contrast when imaged using an energy-integrating detector (Fig. S15). The different materials only become distinguishable once energy-resolved imaging is used. The thinnest multicolor designs with 0.1 mm-thick layers were selected for imaging the phantom. Simulations were also done for the 0.3-mm thick ZnSe:Te and NaI:Tl scintillators for comparison. In all cases, the postprocessing algorithm was used for energy reconstruction.

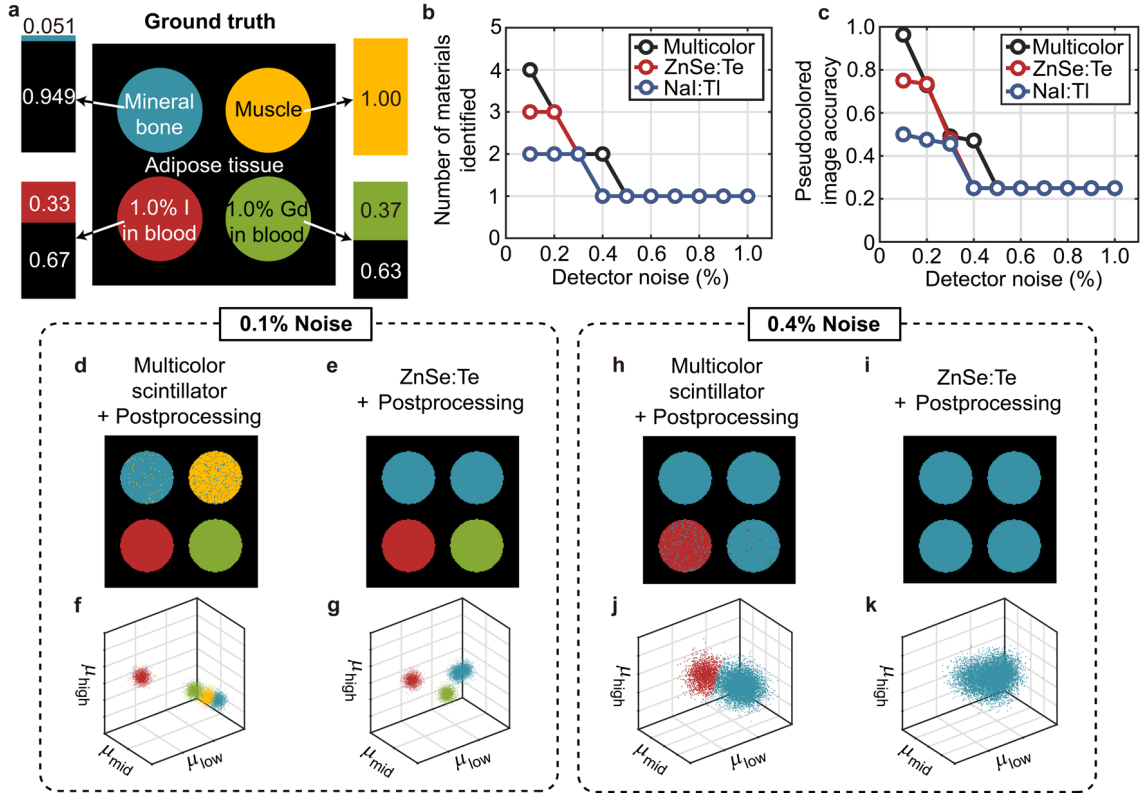

**Fig. 4 Medical phantom simulation results. a** The ground truth phantom with pseudocoloring. The phantom is 1 cm thick and is non-uniform along its depth. The bar plots to the sides of the image show the proportions of the total thickness taken up by each medium versus the adipose tissue. **b** The number of different materials (excluding the background) that were identified using the multicolor, ZnSe:Te, and NaI:Tl scintillators plotted as a function of the detector noise. **c** Accuracy of the pseudocolored image. **d-e** Pseudo-colored phantoms reconstructed using the multicolor and ZnSe:Te scintillators under 0.1% detector noise. **f-g** Energy maps for the multicolor and ZnSe:Te scintillators under 0.1% detector noise. **h-i** Pseudo-colored phantoms reconstructed using the multicolor and ZnSe:Te scintillators under 0.4% detector noise. **j-k** Energy maps for the multicolor and ZnSe:Te scintillators under 0.4% detector noise.

The energy-resolving action of the multicolor scintillator and the postprocessing algorithm can be encompassed by the function $p_{\text{reconstr}}(E_{\text{inc}}, E_{\text{det}})$ calculated with Monte Carlo simulations, which maps the incident intensity spectrum to the reconstructed spectrum. Additive Gaussian noise with a standard deviation proportional to the average image intensity was added to the resulting images for each energy bin[17]. For material identification the effective attenuation coefficient for each energy bin was computed for each pixel. The resulting values were mapped pixel-wise in three-dimensional space and clustered using the same modified *k*-means clustering algorithm used for energy reconstruction postprocessing. This is conceptually similar to the energy maps used in dual-source CT postprocessing[9].

Fig. 4b shows the number of materials (excluding the background) identified as a function of detector noise, where the multicolor design was able to identify the most numbers of materials for all detector noise levels. In order to quantify the accuracy of the reconstructed pseudocolored images, we defined an accuracy metric based on a confusion matrix **C**, where the $ij^{th}$ element corresponds to the number of pixels with integer label *i* in the ground truth image and label *j* in the reconstructed image. The labels of the reconstructed image are then permuted to maximize the diagonal elements of **C**. The accuracy can then be computed as the number of correctly classified pixels (tr{**C**}) divided by the total number of pixels ($N_{\text{pixel}}$):

$$\eta_{\text{img}} = \frac{\text{tr}\{\mathbf{C}\}}{N_{\text{pixel}}} \qquad (8)$$

The multicolor scintillator again demonstrates superior performance compared to the single-layer scintillators by this metric (Fig. 4c). As shown in Fig. 4f-g, the attenuation coefficient cluster centroids are farther apart for the multicolor scintillator due to its superior energy reconstruction accuracy, which allows the *k*-means algorithm to differentiate the mineral bone region from muscle. Material identification becomes more difficult with increasing noise due to the expanding cluster sizes, but the multicolor scintillator maintains the ability to identify the regions with the contrast agents much longer than the single-color scintillators (Fig. 4j-k).

## 2.4 Nanophotonic Filter for Clustering Accuracy Improvement

As described in Section 2.2, the spatial and energy resolution of the multicolor scintillator can be degraded at high X-ray flux densities, as quantified by the FM index. This is primarily caused by the spread of scintillation photons while they propagate toward the detector, which results in overlap between the measured spots – a universal limitation encountered in bulk scintillator materials[6]. Nanophotonics provides us with a method for reducing the spot size on the detector: a nanophotonic angular filter, which we place below the scintillators to reflect photons that travel at high angles.

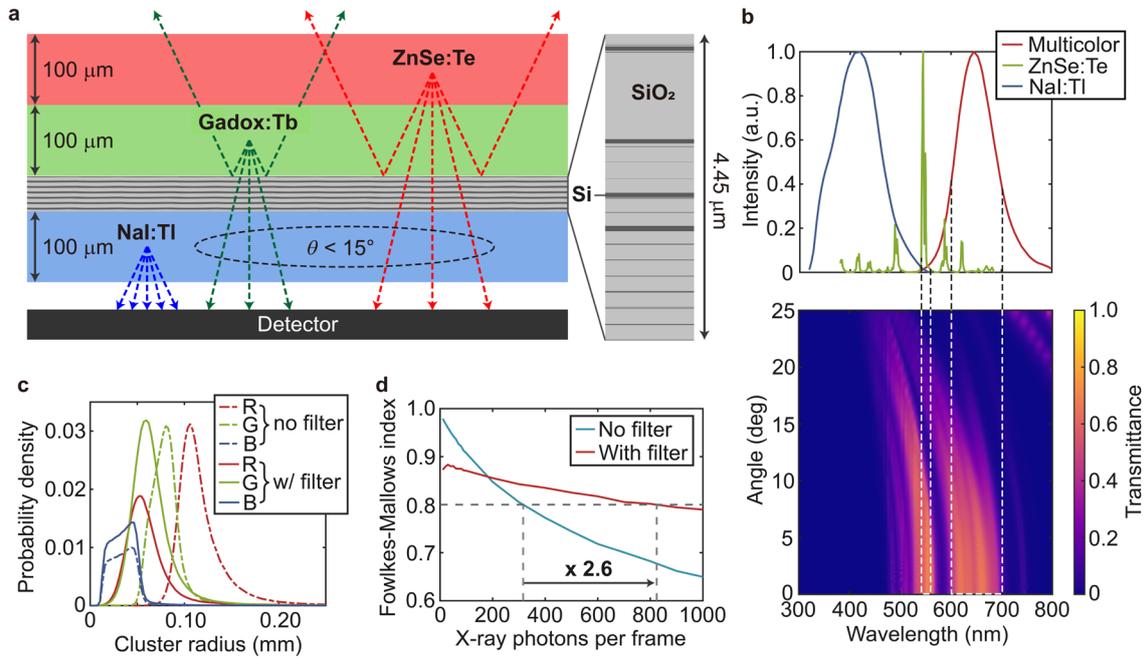

**Fig. 5 X-ray photon flux improvement with a nanophotonic angular filter. a** A schematic showing large-angle filtering by the multilayer filter. **b** Max-normalized emission spectra of the multicolor scintillator constituents (top) and the unpolarized transmittance of the multilayer filter (incidence from Gadox:Tb) as a function of the incidence angle and wavelength. **c** Probability density distributions of the measured cluster radius with and without the multilayer filter. **d** Fowlkes-Mallows index with and without the multilayer filter.

To this end, we designed a multilayer stack of alternating Si and SiO$_2$ layers to reflect any photons incident from above propagating at $\theta \geq 15°$ (Fig. 5a). $\theta = 15°$ was chosen as the threshold angle, because smaller angles filter out ≈ 64% of the scintillation photons that would otherwise reach the detector, which causes the identification of clusters on the detector to be highly inaccurate. Silicon was chosen due to its large refractive index contrast with $SiO_2$, but its absorption in the UV-blue region necessitates the placement of the angle filter to be above NaI:Tl, which is a blue emitter. Details on the optimized filter geometry may be found in Table S2. The optimized filter transmittance for unpolarized topside incidence is shown in Fig. 5b as a function of the incident angle and wavelength. Optimization was done for target wavelengths of 540-560 nm and 600-700 nm which corresponds to the emission wavelengths of Gadox:Tb and ZnSe:Te, respectively. Note that angles above 25° were not considered, because photons propagating at such angles will undergo total internal reflection.

Once the angle filter is inserted in the multicolor scintillator, the overall cluster radii distribution shifts to smaller values for scintillation emission from the top two scintillators (Fig. 5c). This leads to a more gradual decrease of the FM index with the X-ray photon flux as shown in Fig. 5d. Taking FM = 0.8 as the reference value, the corresponding X-ray photon flux was increased by a factor of 2.6, which demonstrates the effectiveness of detector spot size reduction. This demonstrates another use case of nanophotonic designs in enhancing scintillators' performance. Unlike nanophotonic scintillators, our approach does not require the nanoscale patterning of scintillator materials and can leverage existing fabrication methods for multilayer thin films[26,27]. We note that there exists a tradeoff between cluster radius reduction and energy reconstruction accuracy due to the crucial role of a wide cluster radius distribution in our postprocessing method (see Supplementary Information, Section S8).

# 3. Discussion

We have demonstrated that multicolor scintillators can meaningfully improve the energy reconstruction accuracy and contrast agent identification in X-ray imaging. Our multilayer scintillators are designed in conjunction with a dedicated postprocessing method specifically designed to maximize energy reconstruction performance. Compared to previous work on dual-scintillator approaches[14,15], the design rules we have outlined here address the importance of the choice of the scintillating material for each layer beyond their emission colors. In particular, the presence or lack of absorption K-edges in each layer can significantly affect energy reconstruction and material composition analysis, especially when imaging contrast agents. In this respect, adding more scintillators with various K-edge energies within the target energy range to the multilayer stack and subdividing the target energy range into smaller energy bins can provide more finely resolved energy information, which may aid material identification. All scintillators in the stack should still have different emission colors such that their emission spots are distinguishable and to minimize self-absorption.

In Section 2.4, we outlined a method of incorporating a nanophotonic filter to the multicolor scintillator in order to improve the X-ray flux that our framework can handle without energy reconstruction performance degradation. One limitation of the filter is that it causes the loss of scintillation photons that would otherwise have reached the detector. This may lead to larger statistical variation of the cluster size for a given emission depth and noisier images in general due to the low measured scintillation intensity. A possible approach to prevent photon loss while also reducing spot size is to use supercollimation in photonic crystals[28]. The multicolor scintillator can be directly patterned in-plane by etching periodic holes. With the appropriate photonic crystal geometry, the equi-frequency contour can be altered such that scintillation emission is directional along the vertical axis and no longer isotropic.

The multicolor scintillator we have presented uses existing materials which have been put together in macroscopic length scales ($\geq 100$ μm). The nanophotonic filter can also be fabricated relatively easily using physical vapor deposition techniques or even polymer coextrusion depending on the materials used[26,27]. These factors make the multicolor scintillator readily available for use in clinical settings without the need for high-precision, high-cost fabrication procedures. The requirement for RGB detectors may pose some issues, because the Bayer pixel arrangement causes their sensitivity to be up to 3× lower than that of monochrome detectors commonly used in medical imaging. One approach to mitigate this problem is to use nanophotonic patterning, which has been shown to improve scintillator brightness up to 10-fold[5].

We also expect our method to find applications in other areas of scintillation imaging, such as positron emission tomography, where stacked scintillator materials[29,30], in conjunction with pulse analysis algorithms[31], have been proposed to improve coincidence time resolution. Synergetic design in stacked scintillators with reconstruction algorithms may lead to further improvements in coincidence time resolution, towards much anticipated developments in real-time volumetric and millimetric radiopharmaceutical imaging[32].

# 4. Materials and methods

## 4.1 X-ray simulations

X-ray simulations were performed using GEANT4, an ab initio Monte Carlo software for interactions between high-energy particles and matter[18]. Scintillator parameters used in the simulations including scintillation yield and decay time are listed in Table 1 along with references for the refractive index, emission spectra and mass attenuation coefficient data[33-44]. The elemental compositions of human body tissues were taken from the ICRP110 adult male voxel computational phantom[45]. The mass attenuation coefficients of each tissue type were computed from the elemental mass attenuation coefficients provided by the National Institute of Standards and Technology (NIST) using the mixture rule (Eq. (9))[46]:

$$\frac{\mu}{\rho} = \sum_i \omega_i \left(\frac{\mu}{\rho}\right)_i \qquad (9)$$

where $\mu$ and $\rho$ are the linear attenuation coefficient and density, respectively. $\omega_i$ is the proportion by weight of element $i$.

| Scintillator | Scintillation Yield (/keV) | Decay Time (ns) | Refractive Index | Emission Spectrum | Mass Attenuation Coefficient |
|---|---|---|---|---|---|
| ZnSe:Te | 55 (ref. 33) | 5e4 (ref. 33) | Amotchkina et al.[34] | Schotanus et al.[33] | Linardatos et al.[35] |
| Gadox:Tb | 60 (ref. 36) | 1e6 (ref. 37) | 2.3 (ref. 38) | Jung et al.[39] | O'Neill et al.[40] |
| NaI:Tl | 38 (ref. 41) | 230 (ref. 37) | Li et al.[42] | Berkeley Nucleonics Corp[43] | Farzanehpoor Alwars et al.[44] |

**Table 1** Material parameters of the constituents of the multicolor scintillator.

## 4.2 Energy reconstruction accuracy bounds

Randomly guessing the energy bin is the method with the worst performance with an accuracy of 33%. For the upper bound, the accuracy could reach 100% if no restrictions are placed on the energy-dependence of the linear attenuation coefficients of the multicolor scintillators (e.g., when the linear attenuation coefficient spectrum follows a rectangular function). Physically, however, X-ray attenuation in the energy range of our interest (16-67 keV) mainly comes from the photoelectric effect, whose attenuation cross-section has an energy-dependence of $\sigma \propto E^a$ where $a < 0$ (ref. 34). Therefore, to obtain a physically meaningful upper bound, we first parameterized the linear attenuation coefficients of a hypothetical multicolor scintillator as follows:

$$\mu = E^a \exp[b + cH(E - E_K)] \qquad (10)$$

where $H(E - E_K)$ is the Heaviside step function shifted by $E_K$, which models K-edges. The K-edge energies were fixed at 33 keV (bottom layer) and 50 keV (middle layer). The parameters $a$, $b$, and $c$ were numerically optimized for each layer (i.e. a total of nine parameters) with the energy accuracy as the figure of merit. Bounds were set on each parameter based on fitted values for real elements and several scintillators (Fig. S10). The thicknesses of each layer were fixed at the same value during optimization and was swept from 0.01 mm to 3.5 mm to obtain the accuracy upper bound in Fig. 3a.

## 4.3 Modified *k*-means clustering algorithm

The simplest form of *k*-means clustering assumes $N$ clusters are present and randomly initializes the corresponding centroids. The algorithm then (1) assigns each point to the closest centroid and (2) recomputes the centroid of each cluster. The two steps repeat until the cluster centroids no longer change. One of the most significant limitations of this method is that the user must know the number of clusters beforehand, which is not true for stochastic processes such as X-ray scintillation.

Hamerly et al. proposed the *G*-means clustering algorithm to remedy this problem[47]. The algorithm initially starts with a single cluster and performs the Anderson-Darling statistical test with the null hypothesis that the data in the cluster are sampled from a Gaussian distribution. If the null hypothesis is rejected, the cluster is split into two with the new centroids determined by principle component analysis. In our implementation of the *G*-means algorithm, we replaced the statistical test with the more general "dip test", where the null hypothesis is that the cluster is unimodal[48].

## 4.4 Nanophotonic filter optimization

We used the OpenFilters software based on the needle optimization method widely used for designing multilayer stacks to design the multilayer filter[49]. The algorithm alternates between shape optimization, where the layer thicknesses are optimized by gradient descent, and needle insertion, where an infinitesimally thin layer is created at an optimal location in the multilayer stack. The steps repeat until needle insertion is unfavorable across the entire multilayer stack. Refractive index data for Si and $SiO_2$ were taken from Palik[50] and Larouche et al.[49], respectively.

## Acknowledgements

This work is supported in part by the DARPA Agreement No. HO0011249049, as well as being also supported in part by the U. S. Army Research Office through the Institute for Soldier Nanotechnologies at MIT, under Collaborative Agreement Number W911NF-23-2-0121. C. R.-C. is supported by a Stanford Science Fellowship.

## Data availability

The data and codes that support the plots within this paper and other findings of this study are available from the corresponding authors upon reasonable request. Correspondence and requests for materials should be addressed to C. R.-C. (chrc@stanford.edu).

## Conflict of interest

N.R., M.S., and C.R.-C. are seeking patent protection for ideas in this work (US Patent Application 18/286,808 and provisional patent application no. 63/257,611). The remaining authors declare no conflicts of interest.

## Contributions

C.R.-C., M.S., N.R. and S.F. conceived the original idea; S.M. developed the theory with inputs from C.R.-C., M.S. and S.V.; S.M. performed the X-ray scintillation and phantom imaging simulations with the aid of S.C.; S.M. optimized the nanophotonic filter with inputs from C.R.-C.; and C.R.-C. and M.S. supervised the project. The manuscript was written by S.M. and C.R.-C. with inputs from all authors.